\documentclass[twocolumn,prl,showpacs]{revtex4}

\usepackage{graphicx}
\usepackage{amssymb}
\usepackage{amsfonts}

\DeclareGraphicsExtensions{.eps}

\begin{document}
\title{Continuous variable entanglement using cold atoms}

\date{\today}

\author{V. Josse}

\author{A. Dantan}

\author{A. Bramati}

\author{M. Pinard}

\author{E. Giacobino}

\affiliation{Laboratoire Kastler Brossel, Universit\'{e} Pierre et
Marie Curie,\\
Case 74, 4 place Jussieu, 75252 Paris Cedex 05, France}

\begin{abstract}
We present an experimental demonstration of both
\textit{quadrature} and \textit{polarization entanglement}
generated via the interaction between a coherent linearly
polarized field and cold atoms in a high finesse optical cavity.
The non linear atom-field interaction produces two squeezed modes
with orthogonal polarizations which are used to generate a pair of
non separable beams, the entanglement of which is demonstrated by
checking the inseparability criterion for continuous variables
recently derived by Duan \textit{et al.} [Phys. Rev. Lett. {\bf
84}, 2722 (2000)] and calculating the \textit{entanglement of
formation} [Giedke \textit{et al.}, Phys. Rev. Lett. \textbf{91},
107901 (2003)].
\end{abstract}


\newcommand{\beq}{\begin{equation}}
\newcommand{\eeq}{\end{equation}}
\newcommand{\beqr}{\begin{eqnarray}}
\newcommand{\eeqr}{\end{eqnarray}}
\newcommand{\lb}[1]{\label{#1}}
\newcommand{\ct}[1]{\cite{#1}}
\newcommand{\bi}[1]{\bibitem{#1}}
\newcommand{\bk}{_{\bf k}}

\maketitle
\newpage

Entanglement in the continuous variable regime has attracted a lot
of attention in the quantum optics and quantum information fields
in connection with quantum teleportation, cryptography, quantum
computing and dense coding \ct{bennett}. Since the first
realization of quadrature entangled beams by Ou \textit{et al.}
\ct{kimble}, various methods, such as $\chi^{(2)}$ process in
optical parametric amplifier (OPA) \ct{zhang}, or Kerr effect in
optical fibers \ct{silberhorn}, have been used to generate non
separable beams. Recently, the concept of \textit{polarization
entanglement}, i.e. entanglement of Stokes operators between two
beams, has been investigated by Korolkova \textit{et al.}
\ct{korolkova}, and first demonstrated experimentally by Bowen
\textit{et al.} \ct{treps} by mixing two squeezed beams issued
from independent OPAs. The Kerr non-linearity of fibers was also
exploited by Gl\"{o}ckl \textit{et al.} to generate a pulsed
source of polarization entanglement \ct{glockl}.

In this paper we show evidence for continuous variable
entanglement generated using the interaction between a linearly
polarized coherent field and a cloud of cold cesium atoms placed
in a high finesse optical cavity. We demonstrate the entanglement
using the inseparability criterion proposed by Duan \textit{et
al.} and Simon \ct{duan}. We generate two kinds of entanglement
with the same system, quadrature entanglement and polarization
entanglement. For this, we use the recently reported generation of
polarization squeezing \ct{josse} in the field that has interacted
with cold atoms; both the mean field mode and the vacuum mode with
orthogonal polarization exiting the cavity can be squeezed. First,
we show how a direct measurement of the quadrature entanglement of
the beam exiting the cavity can be achieved using two balanced
homodyne detections. We then give the form of the covariance
matrix and the associated \textit{entanglement of formation}
(EOF), which, for Gaussian symmetric states, is directly related
to the inseparability criterion value \ct{giedke}. Last, we
produce two non separable beams by mixing two parts of the
previous outgoing beam with a strong field and achieve
polarization entanglement by locking the relative phases
between the strong field and the weak field exiting the cavity.\\

First, let us consider two orthogonally polarized modes $A_a$ and
$A_b$ of the electromagnetic field satisfying the usual bosonic
commutation rules $[A_{\alpha},A_{\beta}^{\dagger}]=\delta_{\alpha
\beta}$. If
$X_{\alpha}(\theta)=(A_{\alpha}^{\dagger}e^{i\theta}+A_{\alpha}e^{-i\theta})$
and $Y_{\alpha}(\theta)=X_{\alpha}(\theta+\pi/2)$ are the usual
quadrature operators (rotated in the Fresnel diagram by angle
$\theta$), $X_a+X_b$ and $Y_a-Y_b$ are the continuous variable
analogous of the EPR-type operators first introduced by Einstein,
Podolsky and Rosen \ct{einstein}. The criterion of \ct{duan} sets
a limit for inseparability on the sum of the operator variances
\beq \mathcal{I}_{a,b}(\theta)=
\frac{1}{2}\left[\Delta^2(X_a+X_b)(\theta)+
\Delta^2(Y_a-Y_b)(\theta)\right]<2 \label{critere} \eeq For states
with Gaussian statistics, $\mathcal{I}_{a,b}<2$ is a sufficient
condition for entanglement and has already been used in several
experiments to demonstrate continuous variable entanglement
\ct{korolkova,treps,glockl}. Moreover, Giedke \textit{et al.}
recently calculated the EOF of Gaussian symmetric states
\ct{giedke} and showed it to be directly related to the amount of
EPR-type correlations given by (\ref{critere}).

\begin{figure}[b]
\includegraphics[scale=0.7]{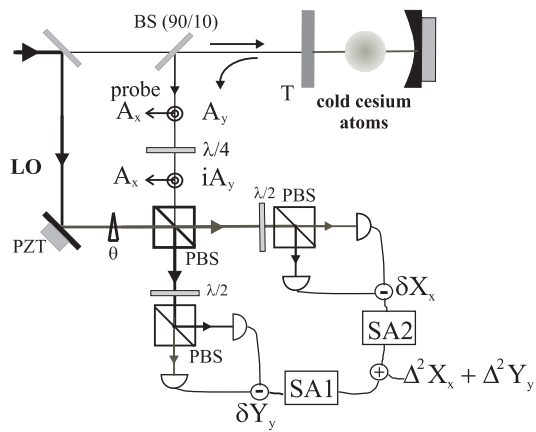}
\caption{Experimental set-up: PBS: polarizing beam splitter;
$\lambda/2$: half-wave plate; PZT: piezo-electric ceramic; SA:
spectrum analyzer. $T=0.1$ is the transmission of the cavity
coupling mirror.}\label{setup}
\end{figure}

In our system, an $x$-polarized beam interacts with a cloud of
cold cesium atoms in an optical cavity. The experimental set-up
\ct{josse} is shown in Fig. \ref{setup}. We probe the atoms with a
linearly polarized laser beam detuned by about 50 MHz in the red
of the 6S$_{1/2}$, F=4 to 6P$_{3/2}$, F=5 transition. The optical
power of the probe beam ranges from 5 to 15 $\mu$W. After exiting
the cavity, both the mean field mode $A_x$ and the orthogonally
polarized vacuum mode $A_y$ are squeezed for frequencies ranging
between 3 and 12 MHz. An interpretation of these results can be
provided by modelling the complicated 6S$_{1/2}$, F=4 to
6P$_{3/2}$, F=5 transition by an X-like four-level atomic
structure \ct{josse1}. When the optical transitions are saturated,
the atoms behave as a Kerr-like medium for the mean field mode
$A_x$, which can be squeezed. Furthermore, the orthogonally
polarized vacuum mode $A_y$ is also squeezed on account of
cross-Kerr effect, but for an \textit{orthogonal} quadrature \ct{josse,josse1}.\\

Our goal is to retrieve the two modes with orthogonal
polarizations which exhibit maximal EPR-type correlations
according to the inseparability criterion (\ref{critere}). We
therefore minimize the quantity $\mathcal{I}_{a,b}(\theta)$ with
respect to $a,b$ and $\theta$. Expanding (\ref{critere}) yields
\beqr \nonumber\mathcal{I}_{a,b}(\theta)&=&\langle \delta
A_a^{\dagger}\delta A_a+\delta A_a\delta A_a^{\dagger}+\delta
A_b\delta A_b^{\dagger}+\delta
A_b^{\dagger}\delta A_b\rangle\\
& & +2[e^{-2i\theta}\langle \delta A_a\delta
A_b\rangle+e^{2i\theta}\langle \delta A_a^{\dagger}\delta
A_b^{\dagger}\rangle] \label{mintheta}\eeqr The right hand side of
the first line in (\ref{mintheta}) is independent of the
polarization basis, while the second line can be written as
$4\cos(2\theta-2\phi) |\langle \delta A_a\delta A_b\rangle|$,
where $2\phi$ is the phase angle of $\langle \delta A_a\delta
A_b\rangle$. Minimizing $\mathcal{I}_{a,b}(\theta)$ corresponds to
maximizing $|\langle \delta A_a\delta A_b\rangle|$ with respect to
$a,b$. In order to find the optimal field components $a,b$, we
introduce another polarization basis $u,v$, such that $\langle
\delta A_u \delta A_v\rangle=0$. It can be shown that there always
exists such a polarization basis and that the $u$ and $v$ modes
quadrature variances are minimal for the same value of $\theta$
\ct{josse3}. The optimal correlations produced in the system are
directly related to the quantum noise properties of these modes
$u,v$. Indeed, the maximally correlated modes $a^*,b^*$ are \beqr
A_{a^*}=(A_u+iA_v)/\sqrt{2},\hspace{0.3cm}
A_{b^*}=(A_u-iA_v)/\sqrt{2}\label{uv}\eeqr and the minimum value
of $\mathcal{I}_{a,b}$ is then given by the sum of the $u,v$ modes
minimal noises \beqr \mathcal{I}_{a^*,b^*}=\min_{a,b,\theta}
\mathcal{I}_{a,b}(\theta)= \min_{\theta} \left[\Delta^2
X_u(\theta)+\Delta^2 X_v(\theta)\right]\label{Iabstar}\eeqr
Consequently, if one or two of the $u,v$ modes are squeezed, the
value $\mathcal{I}_{a^*,b^*}$ corresponding to maximal
correlations is equal to the sum of their squeezing.
Experimentally, one has to look for the $u,v$-type modes, a
signature of which being that $\mathcal{I}_{u,v}(\theta)$ does not
depend on $\theta$ [see (\ref{mintheta})], and measure their
squeezing (if any). The maximally correlated modes are then given
by (\ref{uv}) and the amount of their EPR-type correlations by
(\ref{Iabstar}).

Let us note that modes $u,v$ are not \textit{stricto sensu}
uncorrelated, since $\langle\delta A_u\delta A_v^{\dagger}\rangle$
can be non zero. However, one can think of the correlation
properties of modes $a^*,b^*$ as being created by the mixing of
the $u$ and $v$ modes, as it is usually produced by mixing two
independent squeezed beams \ct{kimble,silberhorn,treps}. Let us
stress that this analysis provides a general framework for finding
out the maximal correlations produced in a two-mode system
exhibiting quantum properties. This method is of interest for a
class of systems such as the optical parametric oscillator in
which the correlations are not produced by mixing independent
beams \ct{fabre}.\\

Coming back to our system, which is symmetrical with respect to
the circularly polarized modes $A_\pm$, it is easy to see that
$\langle \delta A_x \delta A_y\rangle=0$ because $A_x$ and $A_y$
are combinations with equal weights of $A_{\pm}$. Since they are
squeezed for orthogonal quadratures, one can set $A_u=A_x$ and
$A_v=iA_y$, which are now squeezed for the same quadrature. Then,
using (\ref{uv}), the maximally entangled modes are the $\pm
45^{\circ}$ modes to the $x,y$ basis. This gives us the relevant
quantity, $\mathcal{I}_{+45,-45}(\theta)$ , which is to be
measured. Using $A_{\pm 45}=(A_x\pm A_y)/\sqrt{2}$, the
inseparability criterion for the $\pm 45^{\circ}$ modes can be
expressed directly in terms of the $x,y$ modes variances with
$X_u(\theta)=X_x(\theta)$ and $X_v(\theta)=Y_y(\theta)$ \beqr
\mathcal{I}_{+45,-45}(\theta)=\Delta^2 X_x(\theta)+\Delta^2
Y_y(\theta)<2 \label{I45}\eeqr When $\theta$ corresponds to the
angle $\theta_{sq}$ of the squeezed quadrature of $A_x$, both
variances are below unity, and
$\mathcal{I}_{+45,-45}(\theta_{sq})<2$.
\begin{figure}[h]
\includegraphics[scale=0.38]{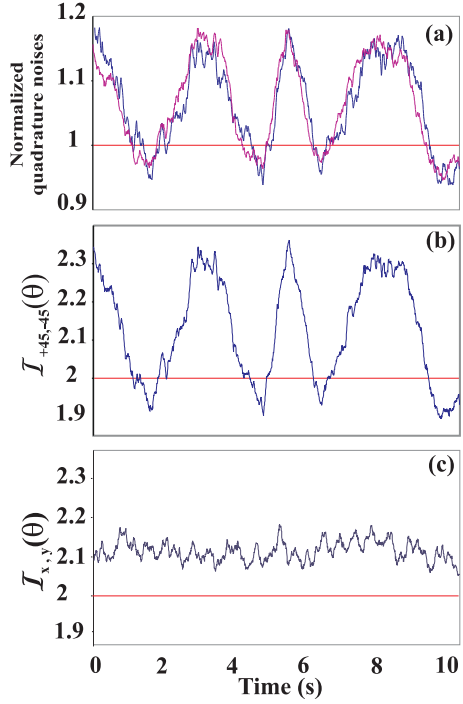}
\caption{(a) Quadrature noise spectra of $A_x$ and $iA_y$, at a
frequency of 5 MHz, when the relative phase $\theta$ between the
LO and the mean field mode is varied in time. (b) Direct
measurement of $\mathcal{I}_{+45,-45}(\theta)$. (c) Corresponding
measurement of $\mathcal{I}_{x,y}(\theta)$.}\label{i45}
\end{figure}

In order to experimentally check the inseparability criterion
(\ref{I45}), we need to simultaneously measure the fluctuations of
$A_x$ and $iA_y$. After the output of the cavity, we insert a
quarter-wave plate that rotates the noise ellipsoid of vacuum mode
$A_y$ by $\pi/2$, the beam is mixed on a beamsplitter with a local
oscillator (LO), and the two resulting beams are sent into two
balanced homodyne detections [Fig. \ref{setup}]. Thus, we
simultaneously measure the quadrature noise of each beam for the
same quadrature. The sum of these two signals directly gives the
sought quantity $\mathcal{I}_{+45,-45}(\theta)$. In Fig.
\ref{i45}(b), we plot a typical measurement of
$\mathcal{I}_{+45,-45}$ as a function of $\theta$, for an analysis
frequency of 5 MHz. Its minimal value is about $1.9$ [1.86
corrected from losses] and corresponds to a case for which $A_x$
and $iA_y$ are both squeezed by about $5\%$ [Fig. \ref{i45}(a)].
\textit{Quadrature entanglement} is thus achieved in a frequency
range given by the cavity bandwidth (3 to 12 MHz).

Consistently with the general method described above, we also
checked that modes $A_x$ and $iA_y$ correspond indeed to
$u,v$-type modes. We therefore measured the quantity
$\mathcal{I}_{x,y}$ in a similar manner as
$\mathcal{I}_{+45,-45}$, and verified that it is independent of
$\theta$ [Fig. \ref{i45}(c)], thus proving that modes $A_{+45}$
and $A_{-45}$ exhibit maximal EPR-type correlations.

Moreover, we note that our measurement not only demonstrates
entanglement, but also quantifies it via the entanglement of
formation. Following Giedke \textit{et al.} \ct{giedke}, we
introduce the covariance matrix (CM) $\gamma$ for the
$\pm45^\circ$ polarized modes: \beqr \nonumber\gamma_{i,j}=\langle
\delta R_i \delta R_j + \delta R_i \delta R_j \rangle/2 \eeqr
where $\{R_i, i=1,...,4\}=\{X_{+45},Y_{+45},X_{-45},Y_{-45}\}$.
Using the fact that $A_x$ and $iA_y$ are uncorrelated and
symmetrical [see Fig. \ref{i45}(b) and \ref{i45}(c)], it is
straightforward to show that the $\pm 45^{\circ}$ modes have
isotropic fluctuations. Choosing $\theta=\theta_{sq}$, the
covariance matrix can be expressed in the standard form given in
Ref. \ct{giedke} \beqr \label{cm}\gamma=\left(
  \begin{array}{cccc}
    n & 0 & k & 0 \\
    0 & n & 0 & -k\\
    k & 0 & n & 0 \\
    0 & -k & 0 & n\\
  \end{array}\right)
\eeqr with $n=\Delta^2 X_{\pm 45}=\Delta^2 Y_{\pm 45}$ and
$k=\langle \delta X_{+45}\delta Y_{-45}\rangle=\langle \delta
X_{-45}\delta Y_{+45}\rangle$ \ct{deux}. As calculated by Giedke
\textit{et al.} the EOF $\mathcal{E}$, representing the amount of
pure state entanglement needed to prepare our entangled state
\ct{bennett2}, is then directly related to the inseparability
criterion value by \ct{giedke} \beqr
\nonumber\mathcal{E}=f(n-k)=f[\mathcal{I}_{+45,-45}(\theta_{sq})/2]
\eeqr with $f(x)=c_+(x)\log_2[c_+(x)]-c_-(x)\log_2[c_-(x)]$ and
$c_{\pm}(x)=(x^{-1/2}\pm x^{1/2})^2/4$. For
$\mathcal{I}_{+45,-45}=1.86\pm 0.02$, the EOF is $\mathcal{E}=0.014\pm 0.003$.\\
\begin{figure}[h]
\includegraphics[scale=0.7]{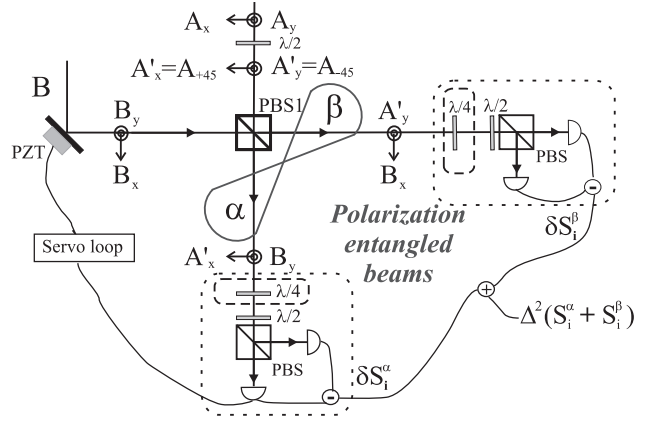}
\caption{Set-up for non separable beams generation. Inserting the
quarter-wave plates (or not) allows for measuring the fluctuations
of $S_3^{\alpha}+S_3^{\beta}$ (or $S_2^{\alpha}+S_2^{\beta}$). The
servo loop is used to lock the B field phase to the squeezed
quadrature angle.}\label{setup2}
\end{figure}

Last, we show that this quadrature entangled beam allows to
generate \textit{polarization entanglement}. Polarization
entanglement for two beams $\alpha$ and $\beta$ \ct{korolkova} is
achieved when \beqr \nonumber\mathcal{I}_{\alpha,\beta}^{S}
=\frac{1}{2}
[\Delta^2(S_2^{\alpha}+S_2^{\beta})+\Delta^2(S_3^{\alpha}+S_3^{\beta})]
 <|\langle S_1^{\alpha}\rangle|+|\langle S_1^{\beta}\rangle|\eeqr
where the $S_i^{\alpha,\beta}$ are the standard quantum Stokes
operators. For this, we produce new modes by mixing the $A_{\pm
45}$ modes studied previously with additional strong fields. The
$A_{\pm 45}$ modes are obtained from the $x,y$ modes by passing
the beam through a half-wave plate with axes at $22,5^{\circ}$.
The fields along the $x$ and $y$ directions are now the $A_{+45}$
and $A_{-45}$ fields, that we will denote by $A'_{x}$ and $A'_{y}$
[see Fig. \ref{setup2}]. The $A'_{x}$ and $A'_{y}$ are then
spatially separated with a polarizing beamsplitter. In the other
input of the beamsplitter, we send a strong field $B$ with a
polarization at $45^{\circ}$ from the beamsplitter axes, yielding
the output fields $B_{y}$ and $B_{x}$. The strong field $B$ is
similar to the local oscillator in the previous experiment, except
that its phase $\theta_B$ is locked to that of one of the $A$
fields by a servo-loop, as shown in Fig. \ref{setup2}. At the two
outputs of the beamsplitter, we have two beams $\alpha,\beta$
which are the superposition of, respectively, $A'_{x}$ and
$B_{y}$, and $A'_{y}$ and $B_{x}$. The Stokes operators
$S_{i}^{\alpha}$ for one of the outputs are then \beqr \nonumber
S_{0}^{\alpha} & = & A'^{\dagger}_{x}A'_{x}+B^{\dagger}_{y}B_{y}
\hspace{0.5cm}S_{1}^{\alpha} = A'^{\dagger}_{x}A'_{x}-B^{\dagger}_{y}B_{y}\\
\nonumber S_{2}^{\alpha} & = &
A'^{\dagger}_{x}B_{y}+B^{\dagger}_{y}A'_{x}\hspace{0.5cm}
S_{3}^{\alpha} = i(B^{\dagger}_{y'}A'_{x}-A'^{\dagger}_{x}B_{y})
\eeqr The Stokes operators $S_{i}^{\beta}$ for the other output
are obtained by exchanging A' and B in the previous expression.
Since the $B$ field is much stronger than the $A$ field, one has
$|\alpha_{B}|\gg |\alpha_{A'}|$, with $\alpha_{B}$ the amplitude
of $B_{x}$ and $B_{y}$ and $\alpha_{A'}$ the amplitude of $A'_{x}$
and $A'_{y}$. Then $\langle S_1^{\alpha}\rangle=-\langle
S_1^{\beta}\rangle\simeq -|\alpha_{B}|^2$ and the inseparability
condition reads
\beqr\mathcal{I}_{\alpha,\beta}^{S}<2|\alpha_{B}|^2\label{Istokes}\eeqr
In this case the Stokes parameters fluctuations are related to
those of the initial $\pm 45^{\circ}$ modes (now denoted
$A'_x,A'_y$) \beqr \delta S_2^{\alpha} & = & \alpha_{B} \delta
X'_{x} (\theta_B) ,\hspace{0.3cm}\;\;\delta S_2^{\beta} =
\alpha_{B} \delta
X'_{y} (\theta_B)\label{s2}\\
\delta S_3^{\alpha}& = &- \alpha_{B} \delta
Y'_{x}(\theta_B),\hspace{0.4cm}\delta S_3^{\beta} = \alpha_{B}
\delta Y'_{y}(\theta_B)\label{s3}\eeqr which straightforwardly
lead to \beqr \nonumber
\mathcal{I}_{\alpha,\beta}^{S}=|\alpha_{B}|^2\hspace{0.1cm}\mathcal{I}_{A'_x,A'_y}
(\theta_B) \equiv|\alpha_{B}|^2\hspace{0.1cm}\mathcal{I}_{+45,-45}
(\theta_B)\eeqr
\begin{figure}[tp]
\includegraphics[scale=0.45]{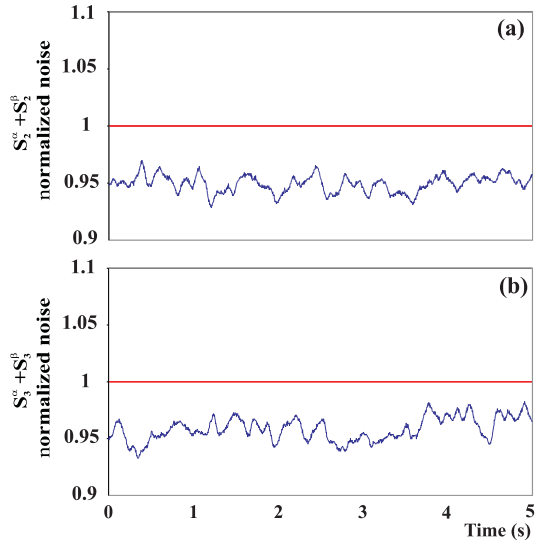}
\caption{Normalized noises at 5 MHz of $S_2^{\alpha}+S_2^{\beta}$
(a) and $S_3^{\alpha}+S_3^{\beta}$ (b), the phase $\theta_B$ being
locked with the value of the squeezed quadrature angle
$\theta_{sq}$.}
\end{figure}\label{pola}
The polarization entanglement condition (\ref{Istokes}) is thus
equivalent to the inseparability criterion (\ref{I45}) for the
$\pm 45^{\circ}$ modes when $\theta_B=\theta_{sq}$. Therefore,
quadrature entanglement can be mapped into a polarization basis
and lead to polarization entanglement \ct{treps}. Experimentally,
we use the set-up shown in Fig. \ref{setup2} and lock the phase of
the B field with the squeezed quadrature angle. We then
successively measure the $S_2$ and $S_3$ Stokes operator noises
using the appropriate combination \cite{korolkova} of plates and
PBS. In Fig. \ref{pola}, we present the normalized quadrature
noises of $S_2^{\alpha}+S_2^{\beta}$ and
$S_3^{\alpha}+S_3^{\beta}$ for an analysis frequency of 5 MHz.
This entanglement between the beams corresponds to a reduction by
approximately $5\%$ in the noise of each variable:
$\mathcal{I}_{\alpha ,\beta}^{S}/|\alpha_{B}|^2=1.9$, consistently
with the quadrature entanglement measurement. From
(\ref{s2}-\ref{s3}), it is also clear that the CM has the same
form as (\ref{cm}).

To conclude, we have reported the generation of continuous
variable entanglement via the interaction with cold atoms in
cavity. First, we have developed a method to directly measure the
inseparability criterion \ct{duan} and demonstrated quadrature
entanglement between two orthogonally polarized modes. The
entanglement was quantified using the entanglement of formation
calculated in Ref. \ct{giedke}. Secondly, we achieve polarization
entanglement after mapping the quadrature entanglement onto two
spatially separated beams.

\begin{acknowledgments}
This work was supported by the QIPC European Project No.
IST-1999-13071 (QUICOV).
\end{acknowledgments}

\end{document}